\documentclass[twocolumn,showpacs,amsmath,amssymb]{revtex4-1}

\usepackage{setspace}
\usepackage{graphicx}
\usepackage{amsmath}
\usepackage{lipsum}
\usepackage{cases}
\usepackage{longtable}
\usepackage{afterpage}
\usepackage{rotating}
\usepackage{threeparttable}
\usepackage{booktabs}
\usepackage{multirow}
\usepackage{subfigure}
\usepackage{array}
\usepackage{color}
\usepackage{float}
\usepackage[section]{placeins}
\usepackage{longtable}
\usepackage[colorlinks=true,linkcolor=red,citecolor=blue,urlcolor=red,]{hyperref}
\usepackage{bm}
\usepackage{lipsum}
\usepackage{appendix}
\usepackage{soul}

\begin{document}

\title{Refined nuclear magnetic octupole moment of $^{113}$In and $^{115}$In}

\author{Fei-Chen Li,$^{1}$ and Yong-Bo Tang,$^{2,*}$}

\affiliation {$^1$Department of Physics, Henan University of Technology, Zhengzhou, 450001, China}
\affiliation {$^2$Physics Teaching and Experiment Center, Shenzhen Technology University, Shenzhen, 518118, China}

\email{tangyongbo@sztu.edu.cn}
\date{\today}
\begin{abstract}
The refined values of the magnetic octupole moments of $^{113}$In and $^{115}$In are obtained by combining high-precision atomic calculations with corresponding hyperfine structure spectrum. We performed an \textit{ab initio }calculations of hyperfine-structure properties for the low-lying states of In atom using the single and double approximated relativistic coupled-cluster method. The hyperfine-structure properties includes first-order hyperfine-structure constants and the second-order magnetic dipole-magnetic dipole, magnetic dipole-electric quadrupole, and electric quadrupole-electric quadrupole effects caused by the off-diagonal hyperfine interaction. Based on our theoretical results, we reanalyze the previously measurements of hyperfine splitting in the 5$p_{3/2}$ state of $^{113}$In and $^{115}$In [Eck and Kusch, Phys. Rev. 106, 958 (1957)], determining corresponding hyperfine-structure constants $A$, $B$, and $C$. By combining these undated HFS constants and our theoretical results, the magnetic octupole moments of $^{113}$In and $^{115}$In nuclei are extracted to be $\Omega(^{113}\rm In)=0.455(44)$~$\mathrm{\mu_{N}\times b}$ , and $\Omega(^{115}\rm In)=0.443(42)$~$\mathrm{\mu_{N}\times b}$, respectively. The refined values of magnetic octupole moments are about smaller 21\% than the previously reported results by Eck and Kusch [Phys. Rev. 106, 958 (1957)]. Additionally,
we also determine the electric quadrupole moment of $^{115}$In nuclei to be $Q(^{115}\rm In)=0.767(9)$ b by combining our theoretical results and the measured values for hyperfine-structure constants of the 5$p_{3/2}$ and 6$p_{3/2}$ states. Our results are compared with available experimental and theoretical results.

\end{abstract}
\maketitle
\section{Introduction}
The nuclear electromagnetic multipole moments are fundamental quantities that describe the shape and electromagnetic distribution of atomic nuclei. It is crucial to have accurate knowledge of these quantities to better understand nucleon-nucleon interactions~\cite{Cocolios_2009_prl,Yordanov_2013_prl,Papuga_2013_prl}. While magnetic dipole moments and electric quadrupole moments are well-known for many nuclei~\cite{Stone_2005_adnd,Pekka_2018_mp}, magnetic octupole moments of many nuclei remain poorly understood. Although it is theoretically possible to evaluate these nuclear multipole moments using nuclear model theory, the accuracy of this approach heavily relies on the specific nuclear model used. A more model-independent alternative to determine the electromagnetic multipole moments of the nucleus is to combine the hyperfine-structure spectrum with corresponding high-precision atomic or molecular calculations. This method is currently one of the most accurate ways to determine the quadrupole moment $Q$ and the octupole moment $\Omega$ of heavy nuclei and unstable nuclei. In fact, the nuclear quadrupole moment $Q$ of many nuclei has been determined accurately using this method~\cite{Pekka_2018_mp}. With advancements in spectroscopic techniques and computational methods, magnetic octupole moments have been determined using this approach for an increasing number of nuclei, such as $^{133}$Cs~\cite{Gerginov_2003_prl,Li_2023_pra}, $^{135,137}$Ba$^{+}$~\cite{Lewty_2013_pra,Sahoo_2013_pra}, $^{87}$Rb~\cite{Gerginov_2009_cjp},and $^{171}$Yb~\cite{Singh_2013_pra,Xiao_2020_pra}.

Indium (Z=49) possesses a proton hole in its nuclear closed shell, making it exhibit rich nuclear properties~\cite{Hinke_2012_nature,Taprogge_2013_prl}. The stable isotopes $^{113}$In and $^{115}$In
, both having considerable high nuclear spin (I=9/2), offer favorable conditions for investigating the effects of high-order hyperfine interactions on hyperfine splitting. In 1957, Eck and Kusch conducted precise measurements of hyperfine splitting in the 5$p_{3/2}$ state of $^{115}$In and $^{113}$In using the conventional atomic-beam techniques~\cite{Eck_1957_pr}. They determined the hyperfine-structure(HFS) constants $A$, $B$, and $C$ of the 5$p_{3/2}$ state through a combination of experimental data and semi-empirical theoretical analysis. They also reported estimated values for the nuclear magnetic octupole moments of $^{115}$In and $^{113}$In, which were approximately half of those predicted by the nuclear single-particle model. In 2009, Gunawardena \textit{et. al.} utilized a two-step, two-color laser spectroscopy technique to measure the hyperfine splitting of the 6$p_{3/2}$ state of $^{115}$In~\cite{Gunawardena_2019_pra}. The corresponding HFS constants $A$, $B$, and $C$ of the 6$p_{3/2}$ state were also determined. Interestingly, the HFS constant $C$ for the 6$p_{3/2}$ state exhibited an opposite sign compared to that of the 5$p_{3/2}$ state. It should be noted that Gunawardena \textit{et. al.} did not consider the correction from second-order effects caused by the off-diagonal hyperfine interaction. Additionally, there have been some theoretical and experimental investigations on the hyperfine structure of indium atoms~\cite{Safronova_2007_pra, Das_2011_jpb,Sahoo_2011_pra,Garcia_2018_prx}, however, most of these studies focused on the magnetic dipole HFS constants.

In this work, our focus is on investigating the nuclear magnetic octupole moments of the $^{115}$In and $^{113}$In. For this purpose, we performed an \textit{ab initio }calculations of first- and second-order HFS constants of $5p_{1/2,3/2}$, $6s_{1/2}$, and $6p_{1/2,3/2}$ states in In atom using relativistic coupled-cluster method at the single and double approximation. Based on our theoretical findings, we reanalyze the experimental results for the $5p_{3/2}$ and $6p_{3/2}$ states~\cite{Eck_1957_pr,Gunawardena_2019_pra}, and extract the corresponding
HFS constants $A$, $B$, and $C$. By combining these undated HFS constants and our theoretical results, we are able to determine the magnetic octupole moments of $^{115}$In and $^{113}$In nuclei. To assess the uncertainty of our results, we also compared the ionization energies, magnetic dipole, and electric quadrupole HFS constants for the $5p_{1/2,3/2}$, $6s_{1/2}$, and $6p_{1/2,3/2}$ states, with available experimental and theoretical values. Detailed numerical results and discussions are presented in the section~\ref{results}.
The following section~\ref{method} provides a brief overview of the hyperfine structure theory, and compiles the HFS expressions for the first-order HFS constants and the second-order corrects caused by the off-diagonal hyperfine interaction. A summary is given in the section~\ref{Summay}.

\section{Theoretical methods}\label{method}

The hyperfine interaction between the nucleus and the electrons causes the fine energy level $E_{J}$ of the atom to split further into hyperfine levels $E_{F}$, $\textbf{F}$=$\textbf{I}$+$\textbf{J}$, where $\textbf{I}$, $\textbf{J}$, and $\textbf{F}$ are the nuclear, atomic, and total angular momentum. We take the 5$p_{3/2}$ and 6$p_{3/2}$ states of $^{115}$In as examples to introduce how to determine the nuclear moment by measuring and calculating the hyperfine structure of atoms.
\begin{figure}[h]
  \centering
  \includegraphics[height=4.6cm,width=8.8cm]{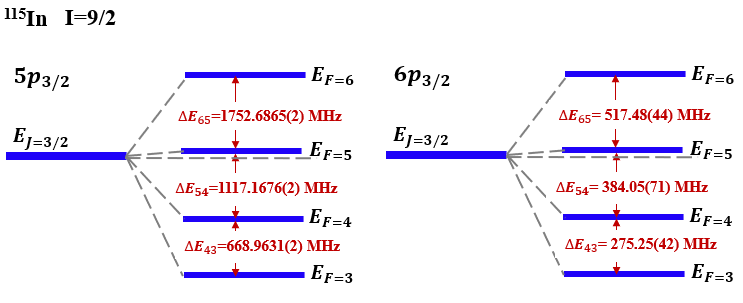}
  \caption{\label{fig1} Schematic diagram of HFS in $^{115}$In for states of 5$p_{3/2}$ and 6$p_{3/2}$, where $\Delta E_{FF^{\prime}}$ denotes the energy difference between two adjacent hyperfine levels determined by experimentd~\cite{Eck_1957_pr,Gunawardena_2019_pra}. }
  \end{figure}
 FIG.~\ref{fig1} shows the hyperfine structure of $^{115}$In of 5$p_{3/2}$ and 6$p_{3/2}$. The hyperfine interval $\Delta E_{FF^{\prime}}=E_{F}-E_{F^{\prime}}$ can be determined experimentally, where $E_{F}$ is the hyperfine level of total quantum number of $F$. When considering second-order HFI,  $E_{F}$ can be expressed as:
  \begin{equation}
		   E_{F}=E_{J}+E^{(1)}_{F}+E^{(2)}_{F},
		 \end{equation}
where $E_{F}^{(1)}$ represents the first-order correction of HFI to the energy, and can be parameterized in terms of the first-order HFS constants such as magnetic dipole ($M1$) HFS constant $A$, electric quadrupole ($E2$) HFS constant $B$, magnetic octupole ($M3$) constant $C$, and etc.. where
\begin{small}
\begin{equation}\label{eq:4}\begin{aligned}
A =\frac{\mu}{I}\frac{\langle {\gamma J}\|T^{(1)}\|{\gamma J}\rangle}{\sqrt{J(J+1)(2J+1)}},
\end{aligned}\end{equation}
\end{small}
\begin{small}
\begin{equation}\label{eq:5}\begin{aligned}
B =2Q \left [\frac{2J(2J-1)}{(2J+1)(2J+2)(2J+3)}\right ]^{1/2}{\langle {\gamma J}\|T^{(2)}\|{\gamma J}\rangle},
\end{aligned}\end{equation}
\end{small}
\begin{small}
\begin{equation}\label{eq:6}\begin{aligned}
C =\Omega \left [\frac{J(2J-1)(J-1)}{(J+1)(J+2)(2J+1)(2J+3)}\right ]^{1/2}{\langle {\gamma J}\|T^{(3)}\|{\gamma J}\rangle}.
\end{aligned}\end{equation}
\end{small}
$E_{F}^{(2)}$ represents the second-order correction of HFI to the energy, and can be parameterized in terms of the second-order HFS constants such as magnetic dipole$-$magnetic dipole ($M1-M1$) HFS constant $\eta$, magnetic dipole$-$electric quadrupole ($M1-E2$) HFS constant $\zeta$, and electric quadrupole$-$electric quadrupole ($E2-E2$) second-order HFS constants $\xi$ and etc.. where
      \begin{small}
      \begin{equation}\label{eq:8}
      \eta=\sum_{J^{\prime}} \frac{(I+1)(2I+1)}{I}{\mu^{2}}\frac{|{\langle {\gamma J^{\prime}}\|T^{(1)}\|{\gamma J}\rangle}|^{2}}{E_{\gamma J}-E_{\gamma J^{\prime}}},
      \end{equation}
      \end{small}
      \begin{small}
      \begin{equation}\label{eq:9}\begin{aligned}
      \zeta=&\sum_{J^{\prime}}\frac{(I+1)(2I+1)}{I}  \sqrt{\frac{2I+3}{2I-1}}\\
            &\times{\mu}Q \frac{{\langle {\gamma J^{\prime}}\|T^{(1)}\|{\gamma J}\rangle}{\langle {\gamma J^{\prime}}\|T^{(2)}\|{\gamma J}\rangle}}{E_{\gamma J}-E_{\gamma J^{\prime}}},
      \end{aligned}\end{equation}
      \end{small}
      \begin{small}
        \begin{equation}\label{eq:10}\begin{aligned}
        \xi=&\sum_{J^{\prime}}\frac{(I+1)(2I+1)(2I+3)}{4I(2I-1)} Q^2 \frac{|{\langle {\gamma J^{\prime}}\|T^{(2)}\|{\gamma J}\rangle}|^{2}}{E_{\gamma J}-E_{\gamma J^{\prime}}}.
        \end{aligned}\end{equation}
        \end{small}
In the Eqs.~(\ref{eq:4})-(\ref{eq:10}), $\left\langle\gamma^{\prime} J^{\prime}\left\|T^{(k)}\right\| \gamma J\right\rangle$ is the reduced matrix element of the spherical tensor operators ~\cite{Beloy_2008_pra} of rank $k$ ($k>0$)  between atomic eigenstate $|\gamma J\rangle$, where $\gamma$ represents the remaining electronic quantum numbers, and $E_{\gamma J}-E_{\gamma J^{\prime}}$ in Eqs.~(\ref{eq:8})-(\ref{eq:10}) represents the interval between two nearby fine-structure levels obtained from Ref.~\cite{NIST_ASD}.
Restricted to $k\leq 3$, the $E_{F}^{(1)}$ can be parameterized in terms of the HFS constants A, B, C.
$\mu$, $Q$ and $\Omega$ are the nuclear magnetic dipole moment, electric quadrupole moment, and magnetic octupole moment, respectively. For $E_{F}^{(2)}$, we focus on $M1$ and $E2$ off-diagonal reduced matrix elements between two nearby fine-structure levels, i.e., $J^{\prime}=J\pm 1$, because these contributions are dominate owing to the small energy denominators.
With these relationships, we can solve for $A$, $B$, and $C$ in terms of the HFS intervals $\Delta E_{FF^{\prime}}$, as well as $\eta$, $\zeta$, and $\xi$. The following expressions are the $A$, $B$, and $C$ constants for the states $np_{3/2}$
\begin{equation}\label{eq1}\begin{aligned}
      A^{np_{3/2}}=&\frac{39}{550}\Delta E_{65}+\frac{64}{825}\Delta E_{54}+\frac{77}{1650}\Delta E_{43}\\
      +&\frac{1}{2970}\eta^{np_{3/2}}-\frac{2}{2475}\sqrt{\frac{2}{5}}\zeta^{np_{3/2}}+\frac{1}{2750}\xi^{np_{3/2}},\\
\end{aligned}\end{equation}
\begin{equation}\begin{aligned}
      B^{np_{3/2}}=&\frac{39}{55}\Delta E_{65}-\frac{16}{55}\Delta E_{54}-\frac{77}{110}\Delta E_{43}\\+&\frac{4}{165}\eta^{np_{3/2}}+\frac{1}{110}\sqrt{\frac{2}{5}}\zeta^{np_{3/2}}+\frac{7}{550}\xi^{np_{3/2}},\\
\end{aligned}\end{equation}
\begin{equation}\label{eq3}\begin{aligned}
      C^{np_{3/2}}=&\frac{21}{1100}\Delta E_{65}-\frac{14}{275}\Delta E_{54}+\frac{7}{200}\Delta E_{43}\\+&\frac{7}{2200}\sqrt{\frac{2}{5}}\zeta^{np_{3/2}}-\frac{7}{11000}\xi^{np_{3/2}}.\\
\end{aligned}\end{equation}
In Eqs.~(\ref{eq1})-(\ref{eq3}), all the required $\Delta E_{FF^{\prime}}$ from Refs.~\cite{Eck_1957_pr,Gunawardena_2019_pra} as shown in the FIG.~\ref{fig1}.
It also can be seen from Eqs.~(\ref{eq1})-(\ref{eq3}) that, in order to accurately extract the first-order HFS constants $A$, $B$, and $C$, the second-order HFS constants $\eta$, $\zeta$, and $\xi$ which contain off-diagonal hyperfine matrix elements, need to be evaluated using atomic structure theory. Subsequently, once we obtain the HFS constants $A$, $B$, and $C$, we can also extract the nuclear moments if the diagonal matrix elements in Eqs.~(\ref{eq:4})-(\ref{eq:6}) are provided. Totally, the diagonal and off-diagonal hyperfine matrix elements are needed to obtain the nuclear moments.

The single particle reduced matrix elements of the operators $T^{(1)}$, $T^{(2)}$, and $T^{(3)}$ are given by
\begin{small}
\begin{eqnarray}\label{eq:88}
\langle{\kappa_{v}}&\|T^{(1)}\|&\kappa_{w}\rangle=-\langle-\kappa_{v}\|C^{(1)}\|\kappa_{w}\rangle
(\kappa_{v}+\kappa_{w})\notag\\
&\times&\int_{0}^{\infty}{dr\frac{P_{v}(r)Q_{w}(r)+P_{w}(r)Q_{v}(r)}{r^2}}\times F(r),
\end{eqnarray}
\end{small}
\begin{small}
\begin{eqnarray}
\langle{\kappa_{v}}\|T^{(2)}\|\kappa_{w}\rangle&=&-\langle\kappa_{v}\|C^{(2)}\|\kappa_{w}\rangle\notag\\
&\times&\int_{0}^{\infty}{dr\frac{P_{v}(r)P_{w}(r)+Q_{v}(r)Q_{w}(r)}{r^3}},
\end{eqnarray}
\end{small}
and
\begin{small}
\begin{eqnarray}\label{eq:99}
\langle{\kappa_{v}}\|T^{(3)}\|\kappa_{w}\rangle&=&-\frac{1}{3}\langle-\kappa_{v}\|C^{(3)}\|\kappa_{w}\rangle
(\kappa_{v}+\kappa_{w})\notag\\
&\times&\int_{0}^{\infty}{dr\frac{P_{v}(r)Q_{w}(r)+P_{w}(r)Q_{v}(r)}{r^4}},
\end{eqnarray}
\end{small}
where the $F(r)$ in Eq.~(\ref{eq:88}) a magnetization distribution model of a finite nucleus, and in this case, we are using a uniform sphere distribution model, and  the relativistic angular-momentum quantum number $\kappa=\ell(\ell+1)-j(j+1)-1/4$, and $P$ and $Q$ are, respectively, the large and small radial components of the Dirac wavefunction. The reduced matrix element
\begin{small}
\begin{eqnarray}\label{eq:99}
\langle\kappa_{v}\|C^{(k)}\|\kappa_{w}\rangle &=&(-1)^{j_{v}+1/2}\sqrt{(2j_{v}+1)(2j_{w}+1)}\notag\\
&\times&\left\{\begin{array}{lll}
      j_{v} & k & j_{w} \\
      1/2 & 0 & -1/2
      \end{array}\right\}\pi(\ell_v,k,\ell_w)
\end{eqnarray}
\end{small}
satisfies the condition $\pi(\ell_v,k,\ell_w)=1$ when $\ell_v+k+\ell_w$ is even, otherwise $\pi(\ell_v,k,\ell_w)=0$.
In this work, We employed a finite basis set, composed of even-tempered Gaussian-type functions expressed as $G_{i}=\mathcal N_{i}r^{\ell+1}e^{-\alpha_{i}r^{2}}$, to expand the Dirac radial wavefunctions $P$ and $Q$ as in Ref.~\cite{Chaudhuri_1999_pra}, where $\mathcal N_{i}$ is the normalization factor, and $\alpha_{i}=\alpha\beta^{i-1}$, with the two independent parameters $\alpha$ and $\beta$ being optimized separately for each orbital symmetries. Table~\ref{basis} lists the Gaussian basis parameters, where $N$ is the size of basis set for each symmetry, and $N_{c}$ and $N_{v}$ represent, respectively, the number of core and virtual orbitals.
 \begin{table}[]
  \newcommand{\RNum}[1]{\uppercase\expandafter{\romannumeral #1\relax}}
  \caption{The parameters of the Gaussian basis set, where $N$ is the size of basis set for each symmetry, and $N_{c}$ and $N_{v}$ represent, respectively,
  the number of core and virtual orbitals. }\label{basis}
\renewcommand\tabcolsep{2.0pt}
\begin{ruledtabular}
\begin {tabular}{lccccccc}
  &$s$&$p$&$d$&$f$&$g$&$h$&$i$
  \\   \hline
  $\eta_{0}\times10^{3}$ &1.5&1&2.5&7.5&15&15&15\\
  $\xi $   &1.92   &1.91   &1.95   &2.0   &2.0   &2.0  &2.0  \\
  $N$      &40    &35     &25     &20     &15    &10   &10   \\
  $N_{c}$  &6      &4      &3      &1      &1     &1    &1    \\
  $N_{v}$  &23     &23     &21     &19    &15    &10   &10   \\
\end{tabular}
\end{ruledtabular}
\end{table}
To accurately calculate the matrix elements in Eqs.~(\ref{eq:4})-(\ref{eq:10}), we need to generate the wave function of the atomic state, which involves solving the electron correlation problem. In the present work, the correlation effects are investigated using $ab$ $initio$ methods at different levels, including the Dirac-Fock (DF) approximation, and linearized and fully single- and double-excitation relativistic coupled-cluster method, denoted respectively by LCCSD and CCSD. The detailed description of our method can be found in previous works for Fr, La$^{2+}$, Ra$^{+}$, Th$^{3+}$, and Cs, in Refs.~\cite{Tang_2017_pra,Lou_2019_aps,Li_2021_jpb,Li_2021,Li_2021_pra,Li_2023_pra}. In practice, the no-pair Dirac Hamiltonian was set as the starting point. The Fermi nuclear distribution was employed to describe the Coulomb potential between electrons and the nucleus.
The virtual orbital with energies smaller than 10000~a.u  and all the core orbital were included in the correlation calculations.

\section{RESULTS AND DISCUSSION}\label{results}

\subsection{Energies}

{
\small
\begin{table*}[ht]\small
\begin{threeparttable}
\newcommand{\RNum}[1]{\uppercase\expandafter{\romannumeral #1\relax}}
\caption{ Energy levels of In I in ${\rm cm^{-1}}$. ${ E_{\rm DF}}$ denotes the lowest-order Dirac-Fock energy.
  ${ E_{\rm LCCSD}}$ and ${ E_{\rm CCSD}}$ are the energies obtained using LCCSD and CCSD approximations, respectively. The values in parentheses shows the relative differences between corresponding calculation results and experimental values ${ E_{\rm Expt.}}$.  }\label{energy1}
\begin{ruledtabular}
\begin {tabular}{lcccccccc}
  Level&${ E_{\rm DF}}$& ${ E_{\rm LCCSD}}$ & ${ E_{\rm CCSD}}$ &${ E_{\rm SD}}$~\cite{Safronova_2007_pra}&${ E_{\rm SDpT}}$~\cite{Safronova_2007_pra}&${ E_{\rm MRCCSD}}$~\cite{Das_2011_jpb}& ${\rm E_{CCSD(T)}}$~\cite{Sahoo_2011_pra2}&${ E_{\rm Expt.}}$\cite{NIST_ASD}
  \\   \hline
  $5p_{1/2}$&	$-$41460(11.2\%)& $-$47006(0.72\%)&	$-$46742(0.15\%)&$-$47061(0.84\%)&$-$46189(1.03\%) &	$-$46804(0.29\%)&	$-$46581(0.19\%)&$-$46670	    \\
  $5p_{3/2}$&	$-$39487(11.2\%)& $-$44860(0.90\%)&$-$44545(0.20\%)&$-$44884(0.96\%)&$-$44031(0.96\%) &	$-$44644(0.42\%)&	$-$44361(0.22\%)&$-$44458   \\
  $6s_{1/2}$&	$-$20567(7.76\%)&	$-$22659(1.62\%)&	$-$22307(0.04\%)            &$-$22668(1.66\%) &$-$22442(0.65\%) &	$-$22539(1.08\%)&	$-$22292(0.02\%)&	$-$22297  \\		
  $6p_{1/2}$&	$-$13972(5.93\%)&	$-$14957(0.70\%)&	$-$14841(0.08\%)            &$-$14943(0.61\%) &$-$14833(0.14\%) &	$-$14896(0.29\%)&	$-$14819(0.23\%)&	$-$14853    \\
  $6p_{3/2}$&	$-$13715(5.77\%)&	$-$14654(0.68\%)&	$-$14545(0.06\%)            &$-$14638(0.57\%) &$-$14532(0.16\%) &	$-$14595(0.28\%)&	$-$14519(0.24\%)&	$-$14555    \\
  $7s_{1/2}$&	$-$9865 (4.85\%)&	$-$10459(0.87\%)&	$-$10370(0.02\%)            &$-$10451(0.80\%) &$-$10381(0.12\%) &	$-$10077(2.81\%)&	                    &	$-$10368    \\

\end{tabular}
\end{ruledtabular}
\end{threeparttable}
\end{table*}
}

We calculated the energies of $5p_{1/2,3/2}$, $6s_{1/2}$, $6p_{1/2,3/2}$, and $7s_{1/2}$ states in In atom using different models including Dirac-Fock(DF),  LCCSD, and CCSD calculations. The predicted energies labeled as ${E_{\rm DF}}$,  ${ E_{\rm LCCSD}}$, and ${ E_{\rm CCSD}}$, respectively, are listed in Table~\ref{energy1}.  These results are compared with available theoretical calculations~\cite{Safronova_2007_pra,Das_2011_jpb,Sahoo_2011_pra2}. The values in parentheses represent the percentage differences between the various calculations and the experimental values from NIST~\cite{NIST_ASD} labeled as ${ E_{\rm Expt.}}$. From Table~\ref{energy1}, one can easily find that: (i) There are noticeable discrepancies between the energies calculated using DF and CCSD methods, indicating significant contributions from electron correlation effects that are not taken into account in DF calculations. The largest deviation occurs at the $5p_{1/2,3/2}$ states,  with an approximate difference of $10\%$.
 (ii) The differences between CCSD results and the experimental values are within $0.2\%$, demonstrating a much better agreement compared to the LCCSD results across all states. This observation suggests that the inclusion of nonlinear terms of the cluster operators is crucial for
 achieving highly accurate energy levels. (iii) It is worth noting that the deviations between the experimental values and our final results are smaller than those of other theoretical calculations, further supporting the validity of our calculation.

\subsection{Hyperfine structure constants $A$}

\begin{table*}[]
  \caption{The HFS constants $A$ (MHz) of $^{115}$In at different correlation levels are given. The ${\rm \delta }$ represents the percentage difference between calculated values and experimental results. The values in brackets are the uncertainties of the recommended values. Some available $ab$ $initio$ theoretical and experimental results are also listed for comparison.  }
  \label{A1}
  \begin{ruledtabular}
  \begin{tabular}{lccccccccccccc}
  Method                     &$5p_{1/2}$     &$5p_{3/2}$  &$6s_{1/2}$    &$6p_{1/2}$&$6p_{3/2}$  &$7s_{1/2}$   \\\hline
  \multicolumn{7}{c}{ Calculated results at different correlation levels}\\
  ${ A_{\rm DF}}$                                &1768           &267         &978           &222     &36       &334   \\
  ${ A_{\rm MBPT(3)}}$                           &2373           &178         &1729          &254     &63       &553   \\
  ${ A_{\rm LCCSD}}$                             &2291           &262         &1810          &260     &82       &561   \\
  ${ A_{\rm CCSD}}$                              &2308           &245         &1689          &261     &71       &527   \\
  \multicolumn{7}{c}{Other theoretical and experimental results}\\
  ${ A_{\rm SD}}$~\cite{Safronova_2007_pra}      &2306           &262.4       &1812          &263.2     &77.82       &544.5    \\
  ${ A_{\rm MRCCSD}}$~\cite{Das_2011_jpb}        &2246           &274         &1736          & 251         &76          & 727        \\
  ${ A_{\rm CCSD}}$~\cite{Sahoo_2011_pra}        &2256(30)       &            &1611(50)      &          &            & 516(30)       \\
  ${ A_{\rm CCSD}}$~\cite{Garcia_2018_prx}       &2260(30)       &257(15)     &1621(50)      &          &            &         \\
  ${ A_{\rm CCSD(T)}}$~\cite{Garcia_2018_prx}    &2274(25)       &253(10)     &1645(37)      &          &            &         \\
  ${ A_{\rm Expt.}}$~\cite{Eck_1957_pr}          &2281.9504(4)   &242.1647(3) &1685.3(6)     &          &   &541.0(3) \\
  ${ A_{\rm Expt.}}$~\cite{Gunawardena_2019_pra} &               &            &              &          &79.33(7)    & \\
  ${ A_{\rm Expt.}}$~\cite{Garcia_2018_prx}  &2282.04(98)    &241.98(51)  &1684.75(1.05) &          &            &          \\
  \multicolumn{7}{c}{Percentage difference between present ${\rm A_{LCCSD}}$, ${\rm A_{CCSD}}$ and the ${\rm A_{Expt.}}$}\\
  ${\rm \delta _{LCCSD}}$  &0.36\%    &8.2\%  &7.0\%  &          & 4.0\%           &  3.2\%        \\
  ${\rm \delta _{CCSD}}$   &1.2\%     &1.1\%  &0.21\% &          &  11\%           &  2.7\%        \\
\end{tabular}
\end{ruledtabular}
\end{table*}
The HFS constants $A$ of $^{115}$In at different correlation levels, including DF, MBPT(3), LCCSD, and full CCSD calculations, are listed in Table~\ref{A1}. The ${\rm \delta_{LCCSD}}$ and ${\rm \delta_{CCSD}}$ represent the percentage differences between the calculated ${ A_{\rm LCCSD}}$, ${ A_{\rm CCSD}}$ values, and the experimental ${ A_{\rm Expt.}}$ value, respectively. In addition, the table includes other theoretical results~\cite{Safronova_2007_pra,Das_2011_jpb,Sahoo_2011_pra,Garcia_2018_prx} and experimental values~\cite{Eck_1957_pr,Gunawardena_2019_pra,Garcia_2018_prx} for comparison with our CCSD results. Table~\ref{A1} demonstrates the significance of considering correlation effects when calculating HFS $A$. For the $6s_{1/2}$, $6p_{3/2}$, and $7s_{1/2}$ states, the total electron correlation effect accounts for almost half of the CCSD results. The inclusion of more comprehensive electron correlation effects leads to results that are closer to the experimental values.
With the exception of $5p_{1/2}$ and $6p_{3/2}$ states, the CCSD results exhibit better agreement with experimental values compared to the LCCSD and MBPT(3) methods. The comparison between ${ A_{\rm LCCSD}}$ and ${ A_{\rm CCSD}}$ reveals that the contributions of nonlinear terms are approximately 1\% for the $5p_{1/2}$, $6p_{1/2}$, and $7s_{1/2}$ states, while they are around 7\% for the $5p_{3/2}$, $6s_{1/2}$, and $6p_{3/2}$ states. This indicates that, even with the same electron correlation effect, the HFS constant of different states within an atom depend on it to varying degrees. We recommend the CCSD values as our final results. By comparing ${ A_{\rm CCSD}}$ with the corresponding ${ A_{\rm Expt.}}$, we can verify the accuracy of our calculation. From the Table~\ref{A1}, it can be observed that ${ \delta_{\rm CCSD}}$ is generally within 3\%, except for the $6p_{3/2}$ state. The difference between the LCCSD result and the experiment for the $6p_{3/2}$ state is about 4\%, significantly smaller than the difference between the CCSD results and the experiment value, which is 11\%. This suggests that higher-order electron correlation effects need to be further considered to obtain more accurate calculation results for this state.

Although the present CCSD, LCCSD~\cite{Safronova_2007_pra}, MRCCSD~\cite{Das_2011_jpb}, CCSD~\cite{Sahoo_2011_pra}, CCSD and CCSD(T)~\cite{Garcia_2018_prx} methods in Table~\ref{A1} are all based on relativistic coupled-cluster-theory, they have some different treatments for electronic correlation effects. Our CCSD results for the $5p_{3/2}$ and $6p_{1/2}$ states exhibit the closest agreement with experimental values when compared to other references listed.  As for the $6p_{1/2}$, $6p_{3/2}$, and $7p_{1/2}$ states, there are no other theoretical values available, but it is evident that our results align well with experimental data for the $6p_{3/2}$ and $7p_{1/2}$ states.

It is important to note that the magnetic dipole moment ($\mu$) used in our calculation of the HFS constants $A$ for $^{115}$In is collected from Ref.~\cite{Flynn_1960} as 5.5408(2)~$\mu_N$. Conversely, employing our theory's $A/\mu$ factor combined with the experimental measurement of ${ A_{\rm Expt.}}$~\cite{Eck_1957_pr,Gunawardena_2019_pra} yields an extracted $\mu$ value of $5.5449~\mu_N$, which is the average for the $5p_{1/2}$, $5p_{3/2}$, $6s_{1/2}$ and $7s_{1/2}$ states. This result agrees with the adopted value of 5.5408(2)~$\mu_N$, indicating not only the effectiveness of our method but also the rationale of obtaining the true nuclear moment value through averaging multiple state results.

\subsection{The electric quadrupole moment $Q$}

\begin{table*}[]
  \begin{threeparttable}
  \newcommand{\RNum}[1]{\uppercase\expandafter{\romannumeral #1\relax}}
  \caption{Determination of the $^{115}$In nuclear electric quadrupole moment using measured $B$ (MHz) and the calculated $B/Q$ from the present work. The present $B/Q$ column contains the results of calculations at different correlation levels from this work as well as those from other\textit{ ab initio} methods in MHz/b. The $Q$ values deduced from other methods are also listed for comparison. $B_{\rm Theo.}$ represents the theoretical value calculated based on our recommended $Q$ and $B/Q$, and $\delta$ represents the percentage difference between $B_{\rm Theo.}$ and $B_{\rm Expt.}$.   }\label{hfsb}
  \begin{ruledtabular}
  \begin{tabular}{lccccrrcrcc}
  \multicolumn{1}{c}{\multirow{2}{*}{\centering $\gamma J$} }
  &\multicolumn{5}{c}{$B/Q$ ($\rm MHz/b$)}
  &\multicolumn{1}{c}{$B_{\rm Expt.}$}
  &\multicolumn{2}{c}{$Q$(b)}
  &\multicolumn{2}{c}{$B_{\rm Theo.}$}
  \\
  \cline{2-6}\cline{8-9}\cline{10-11}
  &\multicolumn{1}{c}{DF}
  &\multicolumn{1}{c}{MBPT(3)}
  &\multicolumn{1}{c}{LCCSD}
  &\multicolumn{1}{c}{CCSD}
  &\multicolumn{1}{c}{Other}
  &\multicolumn{1}{c}{(MHz)}
  &\multicolumn{1}{c}{Present}
  &\multicolumn{1}{c}{Other}
  &\multicolumn{1}{c}{Present}
  &\multicolumn{1}{c}{$\delta$ }
  \\
  \hline
  \multirow{3}{*}{\centering 5$p_{3/2}$ }&415.6   &526.7&590.0  &593.2  &583.5~\cite{Yakobi_2009_CJP} &449.545(3)~\cite{Eck_1957_pr}    &0.758 &0.772(5)~\cite{Eck_1957_pr,Yakobi_2009_CJP}&\multirow{3}{*}{\centering 455 }&\multirow{3}{*}{\centering 1.3\% }\\
            &        &     &       &       &576(4)~\cite{Garcia_2018_prx} &450(1.5)~\cite{Garcia_2018_prx}&0.759&0.781(7)~\cite{Garcia_2018_prx}&&\\
            &        &     &      &     &    &454.2(65)~\cite{Vernon_2022_nature}&0.765    &0.789(13)~\cite{Vernon_2022_nature,Garcia_2018_prx}   &&\\
            \\
  6$p_{3/2}$&55.43   &68.58&84.03 &80.64& &62.5(5)~\cite{Gunawardena_2019_pra}   &0.775& &61.9&1.1\%\\
  \\
  Rec.$Q$     &        &     &      &     &    &                          &$0.767(9)\tnote{a}$& $0.770(8)$~\cite{Stralen_2002_jcp}&&\\
            &        &     &      &     &    &                          &        & 0.76(2)~\cite{Errico_2006_prb}&&\\
            &        &     &      &     &    &                          &        & 0.78(2)~\cite{Errico_2006_prb}&&\\
            &        &     &      &     &    &                          &        & 0.760~\cite{Leiberich_1990_zna}&&\\
            &        &     &      &     &    &                          &        & 0.81~\cite{Belfrage_1984_zpa}&&\\
  \end{tabular}
  \begin{tablenotes}
    \footnotesize
    \item[a] From the average of $Q_{5p_{3/2}}$(=0.758~b) and $Q_{6p_{3/2}}$(=0.775~b).
\end{tablenotes}
  \end{ruledtabular}
  \end{threeparttable}
\end{table*}

Under the same theoretical framework used to calculate the HFS constant $A$, we also calculated the ratio factor $B/Q$ for $5p_{3/2}$ and $6p_{3/2}$ states.
By combining the measured values of $B$ in Refs.~\cite{Eck_1957_pr,Gunawardena_2019_pra}, we extracted the electric quadrupole moment $Q$ of the
the $^{115}$In nuclei. Table~\ref{hfsb} lists the present calculated $B/Q$, as well as the $Q$ derived by combining the experimental $B_{\rm Expt.}$ values, and some other results~\cite{Yakobi_2009_CJP,Garcia_2018_prx,Vernon_2022_nature,Stralen_2002_jcp,Errico_2006_prb,Leiberich_1990_zna,Belfrage_1984_zpa}.

 From Table~\ref{hfsb}, one can see that the $5p_{3/2}$ and $6p_{3/2}$ states exhibit very strong total correlation effects (CCSD-DF), which account for about $30\%$ of the total CCSD results. The nonlinear coupled-cluster terms (CCSD-LCCSD) contribute small correlation effects, approximately 0.5\% and 4.2\% of the total CCSD results for $5p_{3/2}$ and $6p_{3/2}$ states, respectively. The $5p_{3/2}$ state has three experimental $B_{\rm Expt.}$ values, with the highest accuracy being 449.545(3)~MHz reported in 1957~\cite{Eck_1957_pr}. 450(1.5)~MHz reported in 2018 is very close ~\cite{Garcia_2018_prx}. However, the latest reported $B_{\rm Expt.}$ in 2022 is 454.2(65)~MHz~\cite{Vernon_2022_nature}, which is almost 1\% of difference with the previous two values. The $Q(5p_{3/2})$ derived from these three experimental values are 0.758~b, 0.759~b, and 0.765~b, respectively. Since the reported experimental value 449.545(3)~MHz is the most accurate at present, we recommend the $Q(5p_{3/2})$ to be 0.758~b. For the $6p_{3/2}$ state, only one measurement $B_{\rm Expt.}$ has been reported, and when combined with our calculated $B/Q$, a $Q(6p_{3/2})$ of 0.775~b is obtained. Different states have varying sensitivities to correlation effects, and so even when the calculations are performed in the same theoretical framework, there is no guarantee that the results will achieve the same level of accuracy. In this case, taking the average of multiple states may be more reliable. In present work, we take the average of $Q(5p_{3/2})$ and $Q(6p_{3/2})$ as the final $Q$ value of 0.767(9)~b, where the uncertainty in parentheses is the larger one of the two deviations.

 Some other results are also listed in Table~\ref{hfsb} for comparative purposes. In 1984, Belfrage \textit{et al.} obtained a $Q$ value of 0.81~b for $^{115}$In nuclei using the observed hyperfine structure of the atom's 5$s$, 7$p$, and 8$p$ states and an empirically derived $\langle r^{-3}\rangle$ value~\cite{Belfrage_1984_zpa}. Subsequently, a lower value of 0.760~b was proposed based on X-ray data and calculations on the muonic atom~\cite{Leiberich_1990_zna}. Density functional theory calculations on metallic indium yielded $Q$ values of 0.760(20)~b and 0.780(20)~b depending on the density functional used~\cite{Errico_2006_prb}. In 2002, the $Q$ value was determined as 0.770(8)~b by combining the experimental nuclear quadrupole coupling constants and electric field gradients calculated at the four-component CCSD(T) level of theory for four indium halides~\cite{Stralen_2002_jcp}, and it was also considered as a new recommended value of $^{115}$In in the ``year-2008" set of nuclear quadrupole moments updated in 2008~\cite{Pekka_2008_mp}. In 2009, the $B/Q$ value of 583.5~MHz/b~\cite{Yakobi_2009_CJP} calculated by the relativistic Fock-space CCSD method, together with experimental $B_{\rm Expt.}$ of 5$p_{3/2}$ state~\cite{Eck_1957_pr}, yielded a $Q$ value of 0.772(5)~b. This value is very close to the recommended value. In 2018, using the CCSD method, the quadrupole moments $Q$ were extracted as 0.781(7)~b, using the experimental $B_{\rm Expt.}$ of 450(1.5)~MHz~\cite{Garcia_2018_prx},and a calculated $B/Q$ factor of 576(4)~MHz/b from Ref.~\cite{Garcia_2018_prx}. In 2022, the measured $B_{\rm Expt.}$ value of 454.2(65)~MHz~\cite{Vernon_2022_nature} and calculated $B/Q$ from Ref.~\cite{Garcia_2018_prx} yield a $Q$ value of 0.789(13)~b. Both two values are 1.5\% and 2.5\% greater than the recommended value, respectively. Our final $Q$ value of 0.767(9)~b is consistent with the recommended value of 0.770(8)~b. Conversely, when we combine our final $Q$ value of 0.767(9) b with the calculated $B/Q$, the resulted theoretical HFS constant $B$ values for the $5p_{3/2}$ and $6p_{3/2}$ states are 455~MHz and 61.9~MHz, respectively. Our two HFS constant $B$ values have a percentage difference from the experimental values within 1.3\% and 1.1\%, respectively.

\subsection{Magnetic octupole moment}

In this section, we calculated the diagonal and off-diagonal matrix elements for the $5p_{3/2}$ and $6p_{3/2}$ states of $^{115}$In and extracted the magnetic octupole moment.
It is important to note that due to the similarity in mass and nuclear spin of $^{113}$In and $^{115}$In and only slightly different nucleon distribution, the resulted Bohr-Weisskopf (BW) effect has negligible influence on the matrix elements under the current level of accuracy.
 Therefore, the calculated matrix elements can also be used for analyzing the hyperfine interaction of $^{113}$In.

 \begin{table}[]
  \begin{threeparttable}
  \newcommand{\RNum}[1]{\uppercase\expandafter{\romannumeral #1\relax}}
  \caption{ $C/\Omega$ in KHz/($\mu_{N}$$\times$b) and off-diagonal matrix elements in MHz from DF, LCCSD, CCSD calculations. The uncertainty of matrix elements are given in parentheses.}\label{hfsc}
  \begin{ruledtabular}
  \begin {tabular}{ccccc}
  \toprule
    Level&${\rm DF}$&${\rm LCCSD}$ & ${\rm CCSD}$&Final\\   \hline
    \multicolumn{5}{c}{$C/\Omega$ in KHz/($\mu_{N}\times b)$}\\
    $5p_{3/2}$	&	2.312 	&	3.274  	&	3.250 	  &3.250(97)   \\
    $6p_{3/2}$	&	0.309 	&	0.430  	&	0.419 	  &0.419(13)    \\
    \multicolumn{5}{c}{$\text { Off-diagonal matrix elements in MHz }$}\\
    $\langle 5p_{3/2}||{O}^{(1)}|| 5p_{1/2}\rangle$ &$-$249   &$-$452    &$-$507   &$-$507(16)  \\
    $\langle 5p_{3/2}||{O}^{(2)}|| 5p_{1/2}\rangle$ &$-$1039  &$-$1454   &$-$1463  &$-$1463(44) \\

    $\langle 5p_{3/2}||{O}^{(1)}|| 6p_{1/2}\rangle$ &$-$88      &147   &84  &84(13) \\
    $\langle 5p_{3/2}||{O}^{(2)}|| 6p_{1/2}\rangle$ &$-$367  &$-$517   &$-$511  &$-$511(16) \\
    $\langle 5p_{3/2}||{O}^{(1)}|| 6p_{3/2}\rangle$ &$-$306  &$-$638   &$-$566  &$-$566(17) \\
    $\langle 5p_{3/2}||{O}^{(2)}|| 6p_{3/2}\rangle$ &$-$339  &$-$479   &$-$474  &$-$474(15) \\

    $\langle 6p_{3/2}||{O}^{(1)}|| 6p_{1/2}\rangle$ &$-$32    &60        &33       &33(5)       \\
    $\langle 6p_{3/2}||{O}^{(2)}|| 6p_{1/2}\rangle$ &$-$134   &$-$203    &$-$195   &$-$195(6)   \\

  \end{tabular}
  \end{ruledtabular}
  \end{threeparttable}
  \end{table}

Table~\ref{hfsc} presents the hyperfine interaction matrix elements obtained from DF, LCCSD, CCSD calculations. It includes the diagonal $C/\Omega$ in KHz/($\mu_{N}$$\times$b) and important off-diagonal matrix elements in MHz for the $5p_{3/2}$ and $6p_{3/2}$ states.
Regarding the diagonal matrix elements $C/\Omega$ for the $5p_{3/2}$ and $6p_{3/2}$ states, it is observed that the total correlation effects contribute approximately 29\% and 26\% to the total CCSD results, respectively. It is worth mentioning that the electron correlation effects from the nonlinear terms are negative for both states, accounting for around $0.7\%$ and $2.5\%$ of the total CCSD results, respectively.
It is also found that the magnetic dipole off-diagonal matrix elements are more sensitive to the correlation effect than the electric quadrupole off-diagonal matrix elements. For example, the total correlation effect is about $200\%$ for $\langle 5p_{3/2}||{O}^{(1)}|| 6p_{1/2}\rangle$ and $\langle 6p_{3/2}||{O}^{(1)}|| 6p_{1/2}\rangle$.

\begin{table*}[]
  \begin{threeparttable}
    \newcommand{\RNum}[1]{\uppercase\expandafter{\romannumeral #1\relax}}
  \caption{HFS constants $A$, $B$, and $C$ in MHz for the states of 5$p_{3/2}$ and 6$p_{3/2}$ without and with the second-order corrections. $\sum{X_{\Delta E}\times \Delta E}$ are the uncorrected $A$, $B$, and $C$ values, $X_{\eta}\times \eta$, $X_{\zeta}\times\zeta$, and $X_{\xi}\times\xi$ are the second-order corrections due to the $M1$-$M1$, $M1$-$E2$, and $M1$-$E2$ HFI, respectively. The present final results are listed in the ``Total'' column, and the absolute difference between the present results and the other reported results are listed in the last column. $[y]$ denotes the power of 10: $10^y$. }\label{5}
  \begin{ruledtabular}
  \begin{tabular}{ccrrrrrrr}
    \multicolumn{1}{c}{\multirow{2}{*}{\centering $\gamma J$}}
    &\multicolumn{1}{c}{\multirow{2}{*}{\centering HFS}}
    &\multicolumn{5}{c}{Present}
    &\multicolumn{1}{c}{Other}
    \\
    \cline{3-7}
    &
    &\multicolumn{1}{c}{$\sum{X_{\Delta E}\times \Delta E}$}
    &\multicolumn{1}{c}{$X_{\eta}\times \eta$}
    &\multicolumn{1}{c}{$X_{\zeta}\times\zeta$}
    &\multicolumn{1}{c}{$X_{\xi}\times\xi $}
    &\multicolumn{1}{c}{Total}
    &\multicolumn{1}{c}{Reported~\cite{Eck_1957_pr,Gunawardena_2019_pra} }
    &\multicolumn{1}{c}{Diff.}
    \\
  \hline\\
   \multicolumn{9}{c}{{HFS constants of $^{115}$In}}\\\\
   \multirow{3}{*}{\centering 5$p_{3/2}$ } & $A$ &242.164807(23) &4.43(27)[$-$4] &$-$3.55(21)[$-$4]&3.11(19)[$-$5]&242.164926(75)&242.165057(23)&1.3[$-$4]\\
  &$B$      &449.54568(21)     &3.12(19)[$-$2]     &4.00(24)[$-$3]&1.09(7)[$-$3]&449.5827(27)&449.59656(21)&1.4[$-$2] \\
  &$C$      &0.000100(13)        &0.00             &1.40(9)[$-$3]&$-$5.44(33)[$-$5]&0.00144(11)&0.001702(13)&2.6[$-$4]\\\\
  \multirow{3}{*}{\centering 6$p_{3/2}$ } &$A$      &79.33(7)          &1.54(46)[$-$5]     &2.34(35)[$-$5]&4.17(25)[$-$6]&79.33(7)&79.33(7)&\\
  & $B$      &62.5(5)      &1.11(33)[$-$3]     &2.63(40)[$-$4]&1.46(9)[$-$4]&62.5(5)&62.5(5)&\\
  & $C$      &$-$0.04(4)          &0.00             &$-$9.21(1.38)[$-$5] &$-$7.30(44)[$-$6]&-0.04(4)&$-$0.04(4)&\\\\
  \multicolumn{9}{c}{{HFS constants of $^{113}$In}}\\\\
  \multirow{3}{*}{\centering 5$p_{3/2}$ }
 &$A$      &241.641040(58)  &4.41(26)[$-$4]  &$-$3.51(21)[$-$4] &3.05(18)[$-$5]    &241.64113(11)  &241.641293(58)  &1.63[$-$4]\\
 &$B$      &443.41568(52)   &3.17(19)[$-$2]  &3.95(24)[$-$3]    &1.07(7)[$-$3]     &443.4514(29)   &443.46626(52)   &1.49[$-$2] \\
 &$C$      &0.000151(32)    &0.00            &1.38(9)[$-$3]     &$-$5.33(33)[$-$5] &0.00148(13)    &0.001728(45)    &2.49[$-$4]\\
  \end{tabular}
  \end{ruledtabular}
  \end{threeparttable}
  \end{table*}

Since $A/\mu$ and $C/\Omega$ are both magnetic diagonal matrix elements, their correlation trends may be similar.
The diagonal hyperfine matrix elements of the first-order HFS constants in Eqs.~(\ref{eq:4})-(\ref{eq:6}) and the off-diagonal hyperfine matrix elements of the second-order HFS constants in Eqs.~(\ref{eq:8}) and (\ref{eq:9}) are obtained simultaneously,  thus the matrix elements in the Table~\ref{hfsc} should have similar computational accuracy to $A$ and $B$.
From Table~\ref{A1} and Table~\ref{hfsb}, it can be observed that the differences between the CCSD values and the experimental results are within $3\%$ for the states where the total electron correlation does not exceed 50\% and the nonlinear electron correlation does not exceed 10\%. Therefore, we take the CCSD results as the recommended values for the matrix elements except for $\langle 5p_{3/2}||{O}^{(1)}|| 6p_{1/2}\rangle$ and $\langle 6p_{3/2}||{O}^{(1)}|| 6p_{1/2}\rangle$, and estimate the uncertainty of matrix elements as 3\% of the CCSD result. The  $\langle 5p_{3/2}||{O}^{(1)}|| 6p_{1/2}\rangle$ and $\langle 6p_{3/2}||{O}^{(1)}|| 6p_{1/2}\rangle$ strongly dependents on electron correlation effects. From Table~\ref{A1},
It can be seen that ${\rm \delta _{CCSD}}$ of $5p_{3/2}$ and  ${\rm \delta _{CCSD}}$ of $6p_{3/2}$ are $9\%$ and $11\%$, while the $6p_{1/2}$ state has no experimental value. The $6p_{1/2}$ state is not very sensitive to the electron correlation compared with other states in the Table~\ref{A1}. Therefore, we conservatively estimate the uncertainty of these two off-diagonal matrix elements as $15\%$ of the CCSD result.

 Using the off-diagonal hyperfine matrix elements listed in Table~\ref{hfsc}, we can calculate the second-order HFS constants, $\eta$, $\zeta$ and $\xi $ in Eqs.~(\ref{eq:8})-(\ref{eq:10}). We can then evaluate the second-order effects on the correction of the hyperfine structure constants. Table~\ref{5} displays the HFS constants $A$, $B$, and $C$ in MHz without and with considering the second-order corrections for the 5$p_{3/2}$ and 6$p_{3/2}$ states. $\sum{X_{\Delta E}\times \Delta E}$ represents the uncorrected value of $A$, $B$, and $C$, while $X_{\eta}\times \eta$, $X_{\zeta}\times\zeta$, and $X_{\xi}\times\xi$ are the second-order corrections due to the $M1$-$M1$ HFI, $M1$-$E2$ HFI, and $E2$-$E2$ HFI, respectively. The ``Total'' column provides our final results. We have also included other reported results for comparison, along with the absolute differences, listed in the final column.

From Table~\ref{5}, it is apparent that the second-order corrections on  HFS constants $A$, $B$, and $C$ resulted from the $M1$-$M1$, $M1$-$E2$, and $E2$-$E2$ HFI are gradually decreasing. However, for the 5$p_{3/2}$ state, these corrections cannot be disregarded given the current level of accuracy, especially the $M1$-$M1$ and $M1$-$M2$ HFI. Comparatively, our total results for HFS constants $A$, $B$, and $C$ differ from those reported values in Ref.~\cite{Eck_1957_pr}. Nevertheless, our uncorrected HFS constants coincide with the previously reported values in Ref.~\cite{Eck_1957_pr}, thus the difference stem entirely from the differential evaluation of the second-order corrections. For example, the uncorrected HFS constant $C$ value of 0.000100(13) MHz for $^{115}$In, obtained directly from the measured intervals based on first-order HFI, equals the value presented in Ref.~\cite{Eck_1957_pr}.
 On the other hand, the total second-order correction value of $0.00134(11)$ MHz resulted from the off-diagonal HFI is 16\% smaller than previously reported value of $0.001602(32)$ MHz in Ref.~\cite{Eck_1957_pr}. This discrepancy will have a significant impact on the determination of the nuclear moment $\Omega$. The situation is analogous for the 5$p_{3/2}$ state of $^{113}$In.  Regarding the 6$p_{3/2}$ state, our results align with those reported when second-order corrections were not accounted for~\cite{Gunawardena_2019_pra}, as these effects do not manifest at the current level of experimental precision. However, the second-order effects will become relevant when the experimental accuracy surpasses 10~Hz. Our computations can therefore serve as a reference for future, more precise measurements.

    \begin{table}[]
      \begin{threeparttable}
      \newcommand{\RNum}[1]{\uppercase\expandafter{\romannumeral #1\relax}}
      \caption{ The magnetic octupole moment$\Omega$ (in~$\mathrm{\mu_{N}\times b}$) of $^{115}$In and $^{113}$In. The corresponding ${\Omega_{\rm other}}$ from Ref.~\cite{Eck_1957_pr} and the $\Omega_{\rm SP}$ evaluated by the nuclear single-particle model from Ref.~\cite{Xiao_2020_pra} are also listed for comparison. The uncertainty is enclosed in parentheses.}\label{octupole}
      \begin{ruledtabular}
      \begin {tabular}{cccc}
      \toprule
        Isotope&$\Omega_{\rm Present}$&$\Omega_{\rm Other}$~\cite{Eck_1957_pr}&$\Omega_{\rm SP}$~\cite{Xiao_2020_pra}\\
         \hline
        $^{115}$In	&	0.443(42) 	&0.565(12)  	&1.00   \\
        $^{113}$In	&0.455(44) 	&	0.574(15)  	  &0.99    \\
      \end{tabular}
      \end{ruledtabular}
      \end{threeparttable}
      \end{table}

    After determining HFS constant $C$, we can proceed to determine the $\Omega$ by combining the calculated $C/\Omega$ from Table~\ref{hfsc}. For $^{115}$In, the uncertainty of the measured HFS constant $C(6p_{3/2})$ is too large and the sign is abnormal, so the determination of the $\Omega$ is only based on the $C$ of $5p_{3/2}$ state.

    Table~\ref{octupole} displays the present and reported $\Omega$ results (in~$\mathrm{\mu_{N}\times b}$) of $^{115}$In and $^{113}$In.
   The uncertainties are presented in parentheses. The uncertainties of $\Omega_{\rm Present}$ arise from theoretical considerations.
    Observably, the results of the nuclear single-particle mode, $\Omega_{\rm SP}$, far exceed the $\Omega_{\rm Present}$ and $\Omega_{\rm Other}$.
    Specifically, the $\Omega_{\rm other}$ value reported for $^{115}$In in Ref.~\cite{Eck_1957_pr} is 0.565(12)~$\mathrm{\mu_{N}\times b}$.
    By leveraging this reported $\Omega_{\rm other}$ and the HFS constant $C$ value of 0.001702(35)~MHz from Ref.~\cite{Eck_1957_pr},
    we can obtain their $C/\Omega$ value as 3.012~KHz/($\mu_{N}\times b)$. Their $C/\Omega$ value is approximately smaller 7.3\% than our value of 3.250(97)~KHz/($\mu_{N}\times b)$, while their HFS constants $C$ is larger 15\% than our result. We extracted the magnetic octupole moments of $^{113}$In and $^{115}$In nuclei to be $\Omega(^{113}\rm In)=0.455(44)$~$\mathrm{\mu_{N}\times b}$ , and $\Omega(^{115}\rm In)=0.443(42)$~$\mathrm{\mu_{N}\times b}$, respectively. Our refined values are approximately smaller 21\% than the $\Omega_{\rm Other}$ in Ref.~\cite{Eck_1957_pr}. These differences are mainly due to inconsistencies in the evaluation of second-order effects.

\section{Summary}
\label{Summay}
In this work, we used the single and double approximated relativistic coupled-cluster method to
firstly calculate the energies and HFS constants $A$ for $5p_{1/2,3/2}$, $6s_{1/2}$, $6p_{1/2,3/2}$, and $7s_{1/2}$ states in In atom. We also investigated the role of the electron correlation effects in both properties by comparing the results of various approximations with available experimental values. Our results shows that the electron correlation effects, especially the nonlinear corrections of the cluster operators, are very important for precise determinations of these properties. Our CCSD method provides accurate results for both properties. Our CCSD energies agree with experimental values at the level of 0.2\%, while our CCSD HFS constants $A$ differ from the experimental results by no more than $3\%$.

Subsequently, we computed the electric quadrupole hyperfine-structure $B/Q$ factors for the 5$p_{3/2}$ and 6$p_{3/2}$ states of
$^{115}$In, and determined the electric quadrupole moment $Q$ of $^{115}$In nuclei by combining with the measured values for hyperfine-structure constants $B$ of the 5$p_{3/2}$ and 6$p_{3/2}$ states.  Our $Q(^{115}\rm In)$ value of $0.767(9)$ b agrees perfectly with the molecular calculation value of $0.770(8)$ b. We also compared our results with other available theoretical results.

Finally, we conducted an investigation into the second-order effects caused by the off-diagonal hyperfine interaction, namely the magnetic dipole-magnetic dipole, magnetic dipole-electric quadrupole, and electric quadrupole-electric quadrupole effects. These second-order corrections greatly influence the determination of the HFS constant $C$. Utilizing these findings, we reanalyzed the measurements of hyperfine splitting in the 5$p_{3/2}$ and 6$p_{3/2}$ state of $^{115}$In, thereby determining the corresponding HFS constants $A$, $B$, and $C$. Through the combination of these updated HFS constants $C$ and our CCSD result of $C/\Omega$ for the 5$p_{3/2}$ state,  we extracted the magnetic octupole moments of $^{113}$In and $^{115}$In nuclei, which are $\Omega(^{113}\rm In)=0.455(44)$ $\mathrm{\mu_{N}\times b}$ , and $\Omega(^{115}\rm In)=0.443(42)$ $\mathrm{\mu_{N}\times b}$, respectively. Notably, our derived values of $\Omega$ are approximately smaller 21\% than the previously reported results. The present nuclear magnetic octupole moments should be more reliable, and provide a better understanding of nuclear properties of In.

\begin{acknowledgments}
   The work was supported by the National Natural Science Foundation of China under Grant No.12174268 and No.12304269, by the Post-doctoral Research Project of SZTU (Grant No.202028555301011), and the Launching Fund of Henan University of Technology (31401512).
\end{acknowledgments}

%

\end{document}